\documentclass[%
 reprint,
 amsmath,amssymb,
 aps,
]{revtex4-2}

\usepackage{graphicx}
\usepackage{dcolumn}
\usepackage{bm}

\begin{document}

\preprint{}

\title{The solution to the Hardy's paradox}
\thanks{A footnote to the article title}%

\author{Ivan Arraut}
 \altaffiliation[Also at ]{University of Saint Joseph\\
  Estrada Marginal da Ilha Verde, 14-17, Macao, China.}




\date{\today}

\begin{abstract}
By using both, the weak-value formulation as well as the standard probabilistic approach, we analyze the Hardy's experiment introducing a complex and dimensionless parameter ($\epsilon$) which eliminates the assumption of complete annihilation when both, the electron and the positron departing from a common origin, cross the intersection point $P$. We then find that the paradox does not exist for all the possible values taken by the parameter. The apparent paradox only appears when $\epsilon=1$; however, even in this case we can interpret this result as a natural consequence of the fact that the particles can cross the point $P$, but at different times due to a natural consequence of the energy-time uncertainty principle.
\end{abstract}

\maketitle


\section{Introduction}
The Hardy's paradox is a gedanken experiment designed for demonstrating how different is Quantum Mechanics from classical approaches \cite{1}. It is a proof of the non-local character of Quantum Mechanics \cite{2}. The experiment analyzes how an electron and a positron, created initially as a pair (common origin), evolve through different paths. Both particles having then two possible trajectories (two possibilities for each particle) during their evolution. Among all the possible paths, two of them (one for each particle) intersect at a point $P$. In the original derivation of Hardy, total annihilation was assumed for the pair if both particles are able to cross simultaneously the intersection point $P$.\\
The paradox consists precisely in the final detection of patterns which are only possible if both particles travel through the point $P$. Any final detection of patterns is forbidden classically but they occur at the Quantum level, challenging any common sense \cite{1, 2}.\\
A weak-value interpretation of the Hardy's paradox was subsequently done in \cite{3} by using the single and two-particle interpretation of the problem.\\
The single particle approach suggested the necessity for the two particles to cross the intersection point $P$ in order to develop the observed patterns in the instruments. The single particle approach does not care at what time each particle crosses the intersection point. This means that in principle, the particles could cross the point $P$ at different times. 
On the other hand, the two-particle approach showed an apparent contradiction with respect to the single-particle approximation, suggesting the impossibility for the two particles to cross simultaneously the point $P$. Inside the two-particle approach, the results obtained in \cite{3} from the perspective of the weak-value formulation, also suggested that the option for both of the particles to select a path where they do not cross the point $P$, is related to a negative weak-value occupation number. This result was interpreted as a repulsive effect.\\
Although interesting researches in the subject were done \cite{Vaidman, Vaidman2, Vaidman3}, an explanation about the contradictory results between the single and two-particle approximation and the possibility of having a reconciliation between both situations has not been found by the date. In this letter we propose the relevant arguments able to conciliate the single and the two-particle approximations proposed in \cite{3}, as well as the apparent contradictions found inside the probabilistic approximation.\\
We analyze the paradox by introducing a complex and dimensionless parameter $\epsilon$, which allows the possibility for the two particles in the pair to cross the intersection point $P$ without annihilation. \\
In particular, for the singular point $\epsilon=1$, we return back to the Hardy's case where both particles arrive at $P$ and (in principle) annihilate. On the other hand, the case $\epsilon=0$ corresponds to the one where the final desired patters $D^{+-}$ is not detected due to the orthogonality of the initial state and the final desired state. 
Finally, we analyze the cases $\epsilon=2, 4$ and $\epsilon\to\infty$, explaining inside the text why they are so relevant for the analysis. In particular, $\epsilon=4$ corresponds to the case where the detectors never register a $C^{+-}$ event. Interestingly, for this situation, combinations of detection in the form $C^+D^-$ are allowed. In the original Hardy's argument, the possibility of detecting first the electron and after the positron in one frame of reference $F^-$ and then reversing the order of the events by detecting first the positron and after the electron in another frame $F^+$, means that in the laboratory frame of reference, the particles detectors are separated by spacelike intervals. Hardy then assumes that the particles really meet at the point $P$ in all the frames of reference. This is a valid statement, consistent with Special Relativity. However, it is still very restrictive in the sense that there is no guarantee that the particles will really meet at $P$ when they depart from their corresponding origins, even if they travel equal distances. The reason for this is the intrinsic uncertainty of the time arrivals, consistent with the uncertainty principle $\Delta E\Delta t\geq\hbar$. If we take into account the energy-time uncertainty relation, there is no reason for the particles to arrive at the same point $P$ simultaneously and instead they might move slightly separated in the time of arrival at $P$. This is true for most of the possible events consistent with the observed detection patterns. This also remarks that the particles still have some degree of uncertainty in their location at the moment when they are supposed to be at $P$. Then assuming that the particles really meet at $P$ for all the events is a very high restriction, removed in this paper due to the introduction of the parameter $\epsilon$.

\section{The Hardy's paradox: Formulation of the problem}   \label{Sec2}           

\begin{figure}
	\centering
		\includegraphics[width=0.5\textwidth]{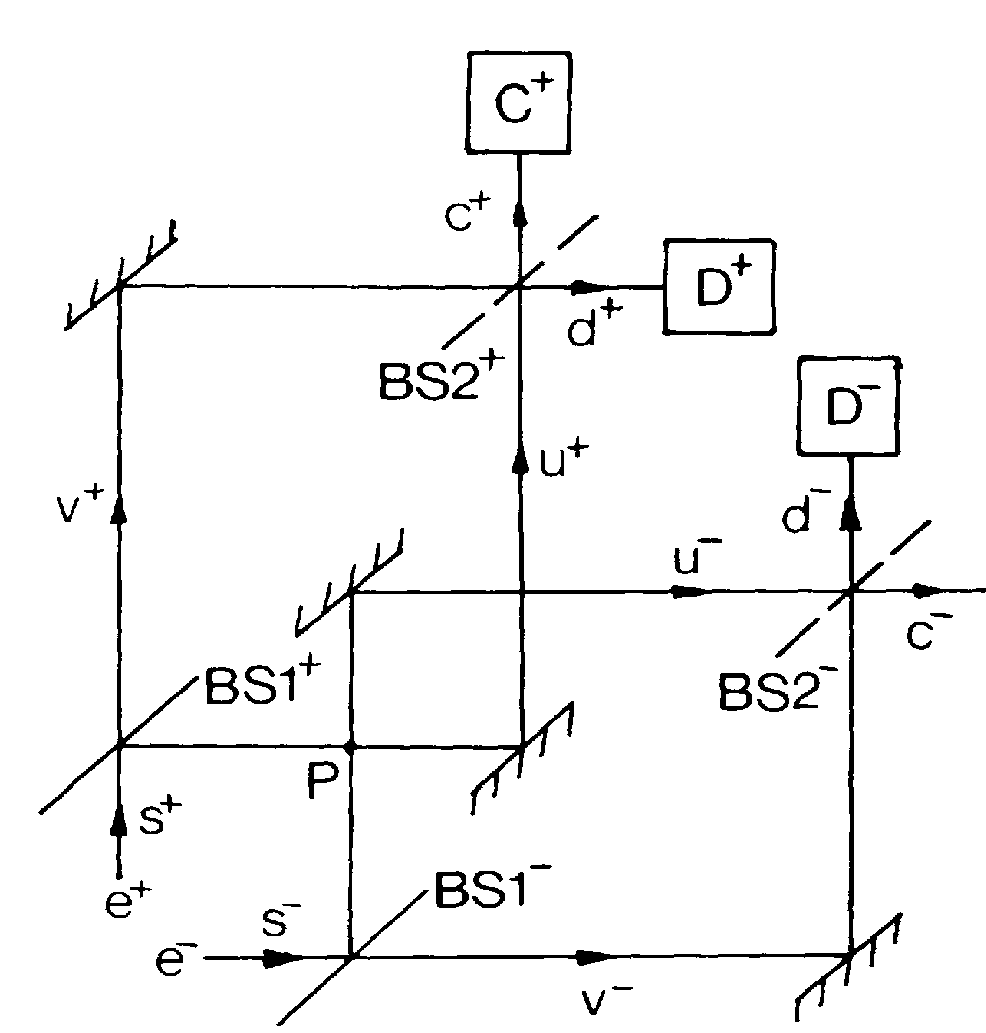}
	\caption{The Hardy’s experiment as it is showed in \cite{1}, illustrating the two possible paths taken by the particles.}
	\label{Fig.1}
\end{figure}

The standard formulation of the Hardy's analysis, starts with the definition of an electron-positron pair emerging from a common point as it is illustrated in the figure (\ref{Fig.1}). There are two possible paths for each particle. Then we have a total of four possibilities. Among all of them, only two have a common intersection point. Hardy's key assumption consists in the complete annihilation of the pairs if they meet at the intersection point $P$. The electron and the positron emerging from the initial common point are represented by the function 

\begin{equation}   \label{eq:1}
\vert\psi>_i=\vert s^+>\vert s^->.
\end{equation}
Subsequently, the electron and the positron take different paths. Once the particles cross the beam splitters defined as $BS^{\pm}_1$ in the figure (\ref{Fig.1}), the functions for the electron and the positron become equivalent to \cite{1}

\begin{equation}   \label{eq:2}
\vert s^{\pm}>\to \frac{1}{\sqrt{2}}\left(i\vert u^\pm>+\vert v^\pm>\right). 
\end{equation}
The effect of the second beam splitter $BS^\pm_2$ can be summarized as

\begin{eqnarray}   \label{eq:3}
\vert u^\pm>\to\frac{1}{\sqrt{2}}\left(\vert c^\pm>+i\vert d^\pm>\right),\nonumber\\
\vert v^\pm>\to\frac{1}{\sqrt{2}}\left(i\vert c^\pm>+\vert d^\pm>\right).
\end{eqnarray}
In \cite{1}, different options related on whether or not the Beam Splitters are removed are analyzed, finding then a contradiction for different cases. The contradiction appears when the final detection of the electron and the positron obey some specific patterns defined as $D^\pm$ in the figure (\ref{Fig.1}). When both beam-splitters are in place, these patterns can only occur if there is interference between the two particles during their journey. However, such interference is only possible if both of the particles are able to cross the point $P$, getting then complete annihilation in agreement with the Hardy's assumption. Although the paradox is valid, the formulation of Hardy is quite restrictive and there are some points to analyze in deeper detail. Hardy's assumption of relative simultaneity for the detection at $D^\pm$ is correct because the interval between the detectors is spacelike. However, assuming that the particles always meet at $P$ for every event is incorrect and restrictive and in disagreement with the uncertainty principle of Quantum Mechanics.\\
Then in general, both particles can cross the intersection point $P$ and still survive the event because in most of the cases one particle will cross first at an instant $t_0$ and the other particle will cross next at an instant $t_0+\Delta t$, with $\Delta t\geq \hbar/\Delta E$. In this letter we will see how we can generalize the Hardy's arguments, solving then any apparent paradox.      

\subsection{Improvements on the Hardy's formulation}

The first assumption which we will change with respect to the Hardy's case is the condition

\begin{equation}   \label{old Hardys}
\vert u^+>\vert u^->=\gamma,
\end{equation}
which is the statement suggesting that when the electron and the positron meet at $P$, they annihilate. In our case, considering the possibility of no-annihilation at $P$, we will take 

\begin{equation}   \label{old Hardys2}
\vert u^+>\vert u^->\to(1-\epsilon)\vert u^+>\vert u^->,\;\;\; near\;\;\; P.
\end{equation}
Here $\epsilon=\vert\epsilon\vert e^{i\alpha}$ is a complex number. The limit $\epsilon\to1$ gives us back the Hardy's result since the photons generated by the annihilation process are assumed not to reproduce pair creation processes \cite{8}. Changing (\ref{old Hardys}) into (\ref{old Hardys2}), modifies the scenario such that the paradox disappears. Here we will show how. 
The definition (\ref{old Hardys2}) excludes absolute annihilation at $P$. A singular case is $\epsilon=1$ where the term $\vert u^+>\vert u^->$ disappears from the scene because in principle, it represents the simultaneous arrival and subsequent annihilation of the electron and the positron when they meet at $P$ in all the frames of reference. After crossing the first Beam Splitters $BS^\pm1$, the state (\ref{eq:1}) becomes

\begin{eqnarray}   \label{acbs}
\vert s^+>\vert s^->\to A(-(1-\epsilon)\vert u^+>\vert u^->+i\vert u^+>\vert v^->\nonumber\\
+i\vert v^+>\vert u^->+\vert v^+>\vert v^->).\;\;\;\;\;
\end{eqnarray}
Here we have used eq. (\ref{eq:2}) inside eq. (\ref{eq:1}) and we have also used the condition (\ref{old Hardys2}). The normalization factor $A$ depends on $\epsilon$ and it is given by 

\begin{equation}
A=\frac{1}{\sqrt{4-(\epsilon+\epsilon^*)+\vert\epsilon\vert^2}}.    
\end{equation}
If we take a frame of reference where the positron has crossed $BS2^+$ but where the electron has not yet arrived at $BS2^-$, then the previous state becomes 

\begin{eqnarray}   \label{thanksgod}
\frac{A}{\sqrt{2}}(-[1-\epsilon]\left[\vert c^+>+i\vert d^+>\right]\vert u^->-\vert c^+>\vert u^->\nonumber\\
+2i\vert c^+>\vert v^->+i\vert d^+>\vert u^->).\;\;\;\;\;
\end{eqnarray}
If the positron is detected at $D^+$, then we project the state (\ref{thanksgod}) on $\vert d^+>$, and the state of the electron is projected to $\frac{A\epsilon}{\sqrt{2}}\vert u^->$ (depending on $\epsilon$). Then we can normalize the event projecting the state (\ref{thanksgod}) toward $\vert u^->$. The we obtain

\begin{equation}   \label{thanksgod2}
\left[U^-\right],\;if\;detection\;at\;D^+,
\end{equation}
with probability 

\begin{equation}   \label{sameprobability}
P_{u^-d^+}=P_{u^+d^-}=\frac{A^2}{2}\vert\epsilon\vert^2.    
\end{equation}
If instead of $F^+$, we now consider the conjugate frame $F^-$, where the electron is detected at $D^-$ before the positron crosses $BS2^+$, by doing an analogous analysis, we get

\begin{equation}   \label{thanksgod3}
\left[U^+\right],\;if\;detection\;at\;D^-,
\end{equation}
with the same probability defined in eq. (\ref{sameprobability}). For this reason we have expressed the equality $P_{u^-d^+}=P_{u^+d^-}$. Note that the probability (\ref{sameprobability}), depends on $\epsilon$ in the same way for both cases, namely (\ref{thanksgod2}) and (\ref{thanksgod3}). This only means that due to the symmetry of the experiment, we can take the frames $F^+$ and $F^-$ to have the same velocity but moving in opposite directions. Then both results must share the same probability. The experiments with an outcome $D^\pm$, occur with a probability

\begin{equation}   \label{thanksgod5}
P_{d^+d^-}=\frac{A^2}{4}\vert\epsilon\vert^2.
\end{equation}
This result can be obtained if we introduce the eq. (\ref{eq:3}) inside eq. (\ref{acbs}) and after projecting over the state $\vert d^+>\vert d^->$ respecting the corresponding normalization factor. Note that the probability of occurrence, depends on $\epsilon$. On the other hand, the projection of the state (\ref{acbs}) over $\vert u^->\vert u^+>$ gives us 

\begin{equation}   \label{thanksgod6}
[U^+U^-],
\end{equation}  
with probability 

\begin{equation}   \label{impressive}
P_{u^+u^-}=A^2(1-(\epsilon+\epsilon^*)+\vert\epsilon\vert^2).    
\end{equation}
In \cite{1, 2}, a reality condition of the form 

\begin{equation}   \label{thanksgod4}
[U^+U^-]=[U^+][U^-],
\end{equation} 
was defined, no matter how we normalize the result $[U^+]$. This relation is defined through the equations (\ref{thanksgod2}), (\ref{thanksgod3}) and (\ref{thanksgod6}). This condition is not completely accurate because the left-hand side in eq. (\ref{thanksgod4}) says that both, the electron and the positron arrive simultaneously at $P$. However, although the right-hand side of the same expression suggests that both, the electron and the positron arrive at $P$, this portion of the equation does not specify if the arrival is simultaneous. Then eq. (\ref{thanksgod4}) is not a precise equality because it does not consider the correction due to the uncertainty on the arrival times on the right-hand side of the expression. This uncertainty is considered by the parameter $\epsilon$.
Then in Hardy's original formulation, the paradox appears from the fact that apparently $[U^+U^-]=0$ during the experiments. 
This is the case because the right-hand side of eq. (\ref{thanksgod4}) corresponds to the case where the particles cross $P$ but not necessarily at the same instant due to the energy-time uncertainty principle $\Delta E \Delta t\geq \hbar$. Mathematically, we can say that while the left-hand side of eq. (\ref{thanksgod4}) is true when $\epsilon=1$, the right-hand side corresponds in general to situations where $\epsilon\neq1$, invalidating then the expression (\ref{thanksgod4}) in general. It is for this reason that we have to revise the Hardy's experiment with the parameter $\epsilon$ included. We could then conclude a connection between $\epsilon$ and the uncertainty in the arrival times at the point $P$.  

\subsection{Relations between probabilities and probability invariants}

\begin{figure}
	\centering
		\includegraphics[width=0.5\textwidth]{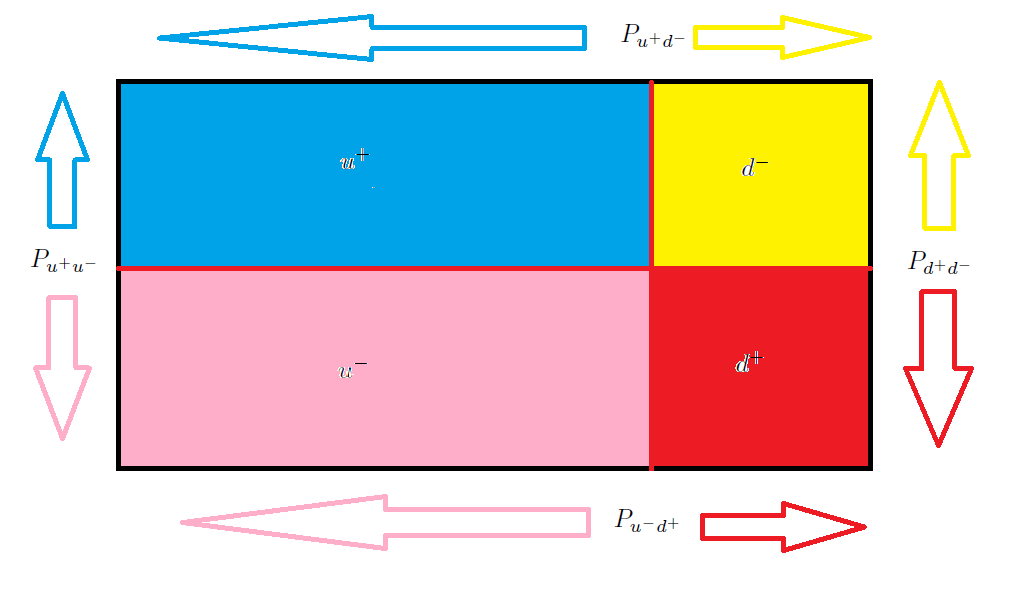}
	\caption{The probability flow through the point $P$. This picture suggests an additional constraint for the probability quantities.}
	\label{Fig.2}
\end{figure}

At this point we can find some relations between probabilities. From eqns. (\ref{sameprobability}) and (\ref{thanksgod5}), it is evident that the following relations are valid

\begin{equation}   \label{Coronaparty1}
P_{u^+d^-}=P_{u^-d^+}=2P_{d^+d^-}=2P_{d^+c^-}=2P_{c^+d^-}.
\end{equation}
From eq. (\ref{impressive}), it is clear that $P_{u^+u^-}$ has a dependence not only on the norm $\vert\epsilon\vert$, but also on the phase $\alpha$, appearing if we expand explicitly the expression as  

\begin{equation}   \label{thelasteq}
P_{u^+u^-}=A^2(1-2\vert\epsilon\vert cos\alpha+\vert\epsilon\vert^2).    
\end{equation}
Then by only knowing the results in eqns. (\ref{sameprobability}) and (\ref{Coronaparty1}), we cannot fix any trustful relation with $P_{u^+u^-}$. In order to find some useful relations between probabilities, we have to evaluate all the other probabilities for the different paths, by using eqns. (\ref{acbs}) and (\ref{thanksgod}), with the corresponding exchanges considering the symmetry of the experiment in eq. (\ref{thanksgod}). The relevant probabilities are

\begin{eqnarray}   \label{Corona2}
P_{u^+v^-}=P_{v^+v^-}=P_{v^+u^-}=A^2,\nonumber\\
P_{c^+v^-}=P_{v^+c^-}=2A^2,\nonumber\\
P_{c^+u^-}=P_{u^+c^-}=\frac{A^2}{2}(4-4\vert\epsilon\vert cos\alpha+\vert\epsilon\vert^2),\nonumber\\
P_{c^+c^-}=A^2\left(4-2\vert\epsilon\vert cos\alpha+\frac{\vert\epsilon\vert^2}{4}\right).
\end{eqnarray}
We can know define the following invariant expressions of probability (independent of $\epsilon$)

\begin{eqnarray}   \label{sing}
P_{v^+v^-}+P_{v^+u^-}+P_{u^+v^-}+P_{u^+u^-}=1,\nonumber\\
P_{d^+d^-}+P_{c^+d^-}+P_{d^+c^-}+P_{c^+c^-}=1,\nonumber\\
P_{u^+d^-}+P_{u^+c^-}+P_{v^+c^-}=1,\nonumber\\
P_{d^+u^-}+P_{c^+u^-}+P_{c^+v^-}=1,
\end{eqnarray}
with the additional condition $P_{v^+d^-}=P_{d^+v^-}=0$. This condition complements the last two equations in (\ref{sing}), which correspond to equations where one particle arrives to the detector before the other one. The consistency of eqns. (\ref{sing}) can be proved with the equations (\ref{acbs}), (\ref{thanksgod}) and other expressions that can be obtained from them and the transformations (\ref{eq:3}). Other two general conditions are 

\begin{eqnarray}   \label{Sing2}
P_{u^+v^-}+P_{u^+u^-}=P_{c^+u^-}+P_{u^+d^-}=\nonumber\\
P_{c^+c^-}+P_{c^+d^-}+P_{u^+d^-}+P_{c^+v^-}.
\end{eqnarray}
The previous expressions cannot constraint the parameter $\epsilon$. However, they mark general invariants that the system must respect. One additional expression, this time being able to constraint $\epsilon$, could be derived if we consider the figure (\ref{Fig.2}).

From this figure, we can write the following relations

\begin{equation}   \label{Constantine}
P_{u^+u^-}+P_{d^+d^-}=P_{u^+d^-}+P_{u^-d^+}.
\end{equation}
If we replace the equations (\ref{sameprobability}), (\ref{thanksgod5}) and (\ref{impressive}), then we get the following quadratic equation

\begin{equation}
\vert\epsilon\vert^2-8\vert\epsilon\vert cos\alpha+4=0. 
\end{equation}
If we solve this equation, then we get

\begin{equation}
\vert\epsilon\vert=4cos\alpha\left(1\pm\sqrt{1-\frac{1}{4cos^2\alpha}}\right).    
\end{equation}
It is clear that given the restriction for $\vert\epsilon\vert$ to be real and positive, then the phase $\alpha$ is restricted to take the range of values

\begin{equation}   \label{allowed1}
\frac{\pi}{3}\geq\alpha\geq -\frac{\pi}{3}.    
\end{equation}
For the extreme cases where $\alpha=\pm\pi/3$, then we get $\vert\epsilon\vert=2$. On the other hand, when $\alpha=0$, we get

\begin{equation}
\vert\epsilon\vert=4\pm2\sqrt{3}.    
\end{equation}
This result suggests that there are two possible values for $\vert\epsilon\vert$ when $\alpha=0$. Then we can say that the allowed values of $\vert\epsilon\vert$ are 

\begin{equation}   \label{allowed2}
4-2\sqrt{3}\geq\vert\epsilon\vert\geq4+2\sqrt{3}.    
\end{equation}
The equations (\ref{allowed1}) and (\ref{allowed2}) represent the allowed values that the phase $\alpha$ and $\vert\epsilon\vert$ can take respectively in each experiment. Note that although the value $\vert\epsilon\vert=1$ appears inside the possible range of values, still this value would correspond to the phase $\alpha=51,3 deg$. Then evidently, the value $\epsilon=1$ is not allowed inside the Hardy's arrangement. Being forbidden the value $\epsilon=1$, then the paradox is solved.  

\subsection{The solution to the Hardy's paradox: Probabilistic approach}   \label{SatisHardy}

The following table summarizes the results obtained from the Hardy's experiment for the different values taken by the parameter $\epsilon$. In agreement with the range of values defined in eqns. (\ref{allowed1}) and (\ref{allowed2}), the values $\epsilon=0$ as well as $\epsilon=1$ and $\epsilon=\infty$ are not allowed, still it is interesting to mention them.   
%
\begin{table} 
\centering
\label{Thetable}
\begin{tabular}{ | m{2em} | m{1cm}| m{2cm} | m{1cm} | m{1cm} | m{2cm}|} 
  \hline
  $\epsilon$ & $P_{d^+d^-}$ & $P_{u^+u^-}$ & $P_{v^+v^-}$ & $P_{c^+u^-}$ & $P_{c^+c^-}$ \\ 
  \hline
  0* & 0 & 1/4 & 1/4 & 1/2 & 1 \\
  \hline
  1* & 1/12 & 0 & 1/3 & 1/6 & 3/4 \\
  \hline
  2 & 1/4 & 1/4 & 1/4 & 0 & 1/4 \\
  \hline
  4 & 1/3 & 3/4 & 1/12 & 1/6 & 0 \\
  \hline
  $\infty*$ & 1/4 & 1 & 0 & 1/2 & 1/4 \\
  \hline
\end{tabular}
\caption{Key probability values for some key values of the parameter $\epsilon$. The asterisk $*$ is put over the values excluded by the allowed ranges defined in eqns.(\ref{allowed1}) and (\ref{allowed2}).
The Hardy's paradox corresponds to the value $\epsilon=1$, which is physically excluded. Note that when $\epsilon\to\infty$ all the events cross the intersection point $P$. However, this value is also physically excluded from the allowed ranges.}
\end{table}
%
From the values obtained in the previous table, it is clear that what is known as the Hardy's paradox in the literature, corresponds to the not allowed value $\epsilon=1$. Note that the previous table deals with the values of $\epsilon$ for which some of the probability values vanish. The remaining probability values can be found from the relations (\ref{Coronaparty1}), (\ref{Corona2}), (\ref{sing}) and (\ref{Sing2}). The range of possible values allowed for $\epsilon=\vert\epsilon\vert e^{i\alpha}$, is obtained from the allowed values defined in eqns. (\ref{allowed1}) and (\ref{allowed2}). The allowed values emerged from the constraint defined in eq. (\ref{Constantine}). The Hardy's paradox then is solved because we have demonstrated that the value $\epsilon=1$ is excluded from the possible values taken by the parameter in agreement with the constraint defined in eq. (\ref{Constantine}). Evidently, without the constraint (\ref{Constantine}), $\epsilon=1$ would just be one among the infinite possibilities taken by the parameter $\epsilon$. Even in such a case, suggesting that $\epsilon$ is exactly would be a huge assumption.    

\section{The weak-value explanation}   \label{Sec.3}

In \cite{3}, an alternative explanation to the Hardy's paradox was done from the perspective of the weak-value. The weak-value is a complex number defined as \cite{5, Tsutsui}

\begin{equation}   \label{eq:4}
X_w=\frac{<\Phi\vert X\vert \psi>}{<\Phi\vert \psi>},
\end{equation}
\\\
where $X$ can be any operator defining an observable. An interesting property of the weak-value is the fact that even if two observables are not compatible, their weak-values can still commute. This is the case because the measurements related to $X_w$ are supposed to be weak enough in order to avoid the limitations related to the uncertainty principle. In eq. (\ref{eq:4}),
$\vert\psi>$ corresponds to an initial state (Pre-selection) and $\vert\Phi>$ corresponds to a final state (Post-selection).\\
In \cite{3}, the wave-functions related to the paths crossing the intersection point $P$, were defined as (Overlapping) $\vert O>_{e, p}$ for the electron and the positron respectively. In the same way, (Non-overlapping) $\vert NO>_{e, p}$, represents the wave-functions corresponding to the paths which never cross the point $P$ for both, the electron and the positron respectively. These states appear after the initial wave-function departing from the lines $s^+$ and $s^-$ in the figure (\ref{Fig.1}) cross the initial Beam Splitters $BS^\pm_1$. The second Beam Splitter defines the Post-selected state in \cite{3}. When the initial state crosses $BS^\pm_1$, the state of the electron-positron pair is defined as

\begin{equation}
\vert\phi>=\frac{1}{4}\left(\vert O>_p+\vert NO>_p\right)\left(\vert O>_e+\vert NO>_e\right).
\end{equation}
In \cite{3}, the Pre-selected state is chosen such that it ignores the contribution $\vert O>_e\vert O>_p$, corresponding to the simultaneous arrival of the electron and the positron to the point $P$. Ignoring this contribution agrees with the Hardy's assumption suggesting that any meeting of the electron and the positron at $P$ is a secure annihilation. Following this argument, we get the following Pre-selected state

\begin{eqnarray}   \label{3}
\vert\psi>=\frac{1}{\sqrt{3}}(\vert NO>_p\vert O>_e+\vert O>_p\vert NO>_e\nonumber\\
+\vert NO>_p\vert NO>_e).
\end{eqnarray}
In the same way as we did before, we will consider later the possibility of including a fraction of the states $(1-\epsilon)\vert O>_e\vert O>_p$ corresponding to the events where both, the electron and the positron can cross the intersection point $P$. The parameter $\epsilon$ will then appear in the analysis when we consider the two-particle cases inside the weak-value formulation. The post-selected state in \cite{3}, is the one corresponding to the case where there is a click for the detectors at $D^+$ and $D^-$ over the figure (\ref{Fig.1}). The Post-selected state is then defined as

\begin{equation}   \label{Post}
\vert\Phi>=\frac{1}{2}\left(\vert NO>_p-\vert O>_p\right)\left(\vert NO>_e-\vert O>_e\right).
\end{equation}
By looking at the single-particle approach, we define the number operators for the electron and positron as

\begin{eqnarray}
\hat{N}^p_{NO}=\vert NO>_p<NO\vert_p,\;\;\;\hat{N}^p_{O}=\vert O>_p<O\vert_p, \nonumber\\
\hat{N}^e_{NO}=\vert NO>_e<NO\vert_e, \;\;\;\hat{N}^e_{O}=\vert O>_e<O\vert_e.
\end{eqnarray}
By introducing these definitions inside eq. (\ref{eq:4}), and by taking the Pre-selected state as (\ref{3})  and the Post-selected state as (\ref{Post}), then we can calculate the weak-value version of the occupation numbers for the single-particle approach as

\begin{eqnarray}   \label{sp}
\hat{N}^e_{O\;w}=1,\;\;\;\;\;\hat{N}^p_{O\;w}=1,\nonumber\\
\hat{N}^e_{NO\;w}=0,\;\;\;\;\;\hat{N}^p_{NO\;w}=0.
\end{eqnarray}
These numbers will be independent of $\epsilon$ even after including the possibility $(1-\epsilon)\vert O>_p\vert O>_e$ inside the pre-selected state (\ref{3}). The result (\ref{sp}) is telling us that for the system to obtain the final desired patterns; both particles (the electron and the positron) must cross the intersection point $P$. The result (\ref{sp}) however, does not specify at what time each particle crosses the intersection point $P$. If we look at the pairs, then we have to work inside a two-particle formalism by defining the weak-value occupation numbers as

\begin{eqnarray}
\hat{N}^{p, e}_{NO, O\;w}=\hat{N}^p_{NO\;w}\hat{N}^e_{O\;w},\;\hat{N}^{p, e}_{O, NO\;w}=\hat{N}^p_{O\;w}\hat{N}^e_{NO\;w},\nonumber\\
\hat{N}^{p, e}_{O, O\;w}=\hat{N}^p_{O\;w}\hat{N}^e_{O\;w},\;\hat{N}^{p, e}_{NO, NO\;w}=\hat{N}^p_{NO\;w}\hat{N}^e_{NO\;w}\;\;\;. 
\end{eqnarray}
By using the same pre-selected and post-selected states, the explicit result for the weak-value version of the pair occupation number is obtained as

\begin{eqnarray}   \label{eq11}
\hat{N}^{p, e}_{NO, O\;w}=1,\;\;\;\;\;\hat{N}^{p, e}_{O, NO\;w}=1, \nonumber\\
\hat{N}^{p, e}_{O, O\;w}=0,\;\;\;\;\;\hat{N}^{p, e}_{NO, NO\;w}=-1.
\end{eqnarray}
These results will have a dependence on $\epsilon$ after introducing this parameter in this formulation. For the moment, we can say that the results obtained in eq. (\ref{eq11}) suggest that in order to get the desired post-selected state $D^\pm$, the electron and the positron must cross the intersection point $P$ (results $\hat{N}^{p, e}_{NO, O\;w}=1$ and $\hat{N}^{p, e}_{O, NO\;w}=1$). However, they cannot cross $P$ simultaneously as the result $\hat{N}^{p, e}_{O, O\;w}=0$ suggests. Indeed, the weak-value number $\hat{N}^{p, e}_{O, O\;w}$, is a number able to specify if the particles cross simultaneously the point $P$ or not. Basically, if $\hat{N}^{p, e}_{O, O\;w}$ vanishes, the number is telling us that no particle can appear at the same time at $P$ and then survive the event.
In other words, the role of $\hat{N}^{p, e}_{O, O\;w}$ is to measure the differences on the arrival times at $P$, between the electron and the positron.\\ 
We must remark once again that the single-particle approach of the weak-value formulation cannot specify whether or not the particles arrive at the same time ({\bf simultaneous}) at $P$ or at different times. All what eq. (\ref{sp}) says is that the particles must cross the point $P$ if we want to get $D^\pm$ on the detectors. For this reason, the two-particle approach formulated in eq. (\ref{eq11}) is very important.\\
At this point we can see that from the perspective of the weak-value approximation, the apparent paradox can be interpreted as an apparent disagreement between the single and the two-particle approximation related to the events happening at $P$ during the evolution of the pair (electron-positron) inside the system. This apparent disagreement comes out from a the standard interpretation of the results related to the two-particles number in eq. (\ref{eq11}) in the original approaches of Hardy and in \cite{3}. In this paper we reinterpret this results by introducing the parameter $\epsilon$. \\ 
Note that there is an intriguing result connected to the event related to the evolution of the particles through paths not crossing $P$. In eq. (\ref{eq11}), $\hat{N}^{p, e}_{NO, NO\;w}=-1$, suggests that these events are related to a negative weak-value occupation number. In \cite{3} this is interpreted as a repulsive effect. This can be also interpreted as a shift of the phases of the particles moving through the system. This means that a negative weak-occupation number can be expressed as $\hat{N}^{p, e}_{NO, NO\;w}=-1=e^{i\pi}$, with a phase difference of $\pi$ between the electron and the positron moving through the system. $\hat{N}^{p, e}_{NO, NO\;w}$ is interpreted in general as a number measuring the events where the particles do not cross the intersection point $P$ simultaneously.  

\section{Improvements of the weak-value approximation: The inclusion of the states $(1-\epsilon)\vert O>_p\vert O>_e$}

We can introduce the parameter $\epsilon$ inside the weak-value formalism. The only change will appear in the pre-selected state defined initially in eq. (\ref{3}). Note that if we include the term $\vert O>_p\vert O>_e$ in eq. (\ref{3}), all the relevant weak-values would diverge since the Pre-selected state would be orthogonal to the Post-selected state defined in eq. (\ref{Post}). The divergence disappears if we introduce the parameter $\epsilon$ in the form $(1-\epsilon)\vert O>_p\vert O>_e$, with $\epsilon\neq0$ in the Pre-selected state (\ref{3}). This variation on the Post-selected state is equivalent to the change done in eq. (\ref{old Hardys2}), when we analyzed the probabilities. In this way we obtain

\begin{eqnarray}   \label{Real pre}
\vert\psi>=A(\vert NO>_p\vert O>_e+\vert O>_p\vert NO>_e\;\;\;\;\;\;\;\;\;\;\;\;\;\nonumber\\
+\vert NO>_p\vert NO>_e+(1-\epsilon)\vert O>_p\vert O>_e).
\end{eqnarray}  
Here $A$ is the same normalization factor used in eqns. (\ref{acbs}) and (\ref{thanksgod}) when we analyzed the original probability formulation. The redefinition of the Pre-selected state does not affect the single-particle results obtained in eq. (\ref{sp}). However, it will affect the two-particle results obtained in eq. (\ref{eq11}). The modifications for the two-particle weak-value numbers are 

\begin{eqnarray}   \label{eq14}
N^{p, e}_{NO, O\;w}=\frac{e^{-i\alpha}}{\vert\epsilon\vert}, \;\;\;\;\;N^{p, e}_{O, NO\;w}=\frac{e^{-i\alpha}}{\vert\epsilon\vert},\nonumber\\
N^{p, e}_{O, O\;w}=1-\frac{e^{-i\alpha}}{\vert\epsilon\vert}, \;\;\;\;\;N^{p, e}_{NO, NO\;w}=-\frac{e^{-i\alpha}}{\vert\epsilon\vert},
\end{eqnarray}
where we have used $\epsilon=\vert\epsilon\vert e^{i\alpha}$. The probability for the detection of the patterns $D^\pm$, can be obtained by projecting the Post-selected state over the Pre-selected state. We obtain in this way

\begin{equation}   \label{conhp}
\vert<\Phi\vert \psi>\vert^2=\frac{A^2\vert\epsilon\vert^2}{4}=\frac{\vert\epsilon\vert^2}{4(4-(\epsilon+\epsilon^*)+\vert\epsilon\vert^2)},
\end{equation}
consistent with the result obtained in eq. (\ref{thanksgod5}). Note that in general, $P_{d^+d^-}$ depends on $\epsilon$. If we choose $\epsilon=1$, then we get $P_{d^+d^-}=1/12$, consistent with the results obtained in \cite{3}. This makes sense because for $\epsilon=1$, we return-back to the Pre-selected state (\ref{3}), which avoids the inclusion of the option $\vert O>_p\vert O>_e$. It can be proved from eq. (\ref{conhp}), that more generally $P_{d^+d^-}=1/12$ if 

\begin{equation}   \label{P112}
\vert\epsilon\vert=-\frac{cos\alpha}{2}\left(1-\frac{1}{\vert cos\alpha\vert}\sqrt{cos^2\alpha+8}\right).
\end{equation}
Then there is a full family of parameters $\epsilon$ for which $P_{d^+d^-}=1/12$. This means that there is nothing special about the value $\epsilon=1$ after all because there are plenty of possibilities such that we can get outputs of the experiment of Hardy for which there is detection at $D^{\pm}$ and still the particles cross the intersection point $P$. 
%
\begin{table}
\centering
\label{Thetable}
\begin{tabular}{ | m{2em} | m{1cm}| m{1.5cm} | m{1.5cm} | m{1.5cm} |} 
  \hline
  $\epsilon$ & $N^{e, p}_{O, Ow}$ & $N^{e, p}_{NO, Ow}$ & $N^{e, p}_{O, NOw}$ & $N^{e, p}_{NO, NOw}$\\ 
  \hline
  0* & $-\infty$ & $\infty$ & $\infty$ & $-\infty$\\
  \hline
  1* & 0 & 1 & 1 & -1\\
  \hline
  2 & 1/2 & 1/2 & 1/2 & -1/2\\
  \hline
  4 & 1/4 & 1/4 & 1/4 & -1/4\\
  \hline
  $\infty$* & 1 & 0 & 0 & 0\\
  \hline
\end{tabular}
\caption{Key values for the weak-value occupation number for the electron-positron pair. They correspond to some key values of the parameter $\epsilon$. The asterisk $*$ over some of the values means that they are excluded from the range of possible values in agreement in eqns. (\ref{allowed1}) and (\ref{allowed2}). The Hardy's paradox corresponds to the value $\epsilon=1$, which is one of the mentioned forbidden values in the system. Note that when $\epsilon\to\infty$ all the events cross the intersection point $P$ and then $N^{p, e}_{O, O\;w}\to1$. However, this is also another forbidden value.}
\end{table}
%
Finally, it is important to remark that the following condition over the pair of particles

\begin{equation}
N^{p, e}_{NO, O\;w}+N^{p, e}_{O, NO\;w}+N^{p, e}_{O, O\;w}+N^{p, e}_{NO, NO\;w}=1,    
\end{equation}
is just equivalent to the equation $P_{v^+v^-}+P_{v^+u^-}+P_{u^+v^-}+P_{u^+u^-}=1$, showed in eq. (\ref{sing}).  

\section{Conclusions}	

In this paper we have found a novel formulation to analyze the Hardy's paradox. We have found that there are many different ways for the electron and the positron to cross the intersection point $P$ without annihilation. We have introduced a complex parameter $\epsilon$, which in general allows the possibility for the particles in the pair to cross $P$ without annihilation. The same parameter conciliates the single and two particle approaches for the weak-value formulation as it has been analyzed within this paper. This conciliation, suggests that the reality condition (\ref{thanksgod4}) is wrong because it corresponds to the equality of a quantity which suggests that the particles arrive at the same time at $P$ (left-hand side), with a quantity which does not care at what time both particles arrive (right-hand side). In fact, the inclusion of the parameter $\epsilon$ allows both particles to cross $P$ but not necessarily at the same time. The fact that both particles not necessarily cross the point $P$ at the same time, is a natural consequence of the energy-time uncertainty principle $\Delta E\Delta t\geq\hbar$. Then even if the two particles depart at the same time, with the same energy; at the moment of measuring the arrival time at $P$, one particle will register the travel interval $t$, while the other will register $t\pm\Delta t$, where $\Delta t$ is consistent with the uncertainty principle. Finally, we must remark that in this paper we demonstrated that the value $\epsilon=1$, which corresponds to the Hardy's paradox value, is forbidden due to the constraint defined in eq. (\ref{Constantine}). This constraint gave us all the possible values for $\vert\epsilon\vert$ and $\alpha$ inside the experiment. Those range of values can be found in eqns. (\ref{allowed1}) and (\ref{allowed2}).    \\\\


\begin{acknowledgments}
The author would like to thank Prof. Izumi Tsutsui for his kind attention during our discussions about these results during the last three years at the KEK High Energy Accelerator Research organization (Theory Center).
\end{acknowledgments}



\end{document}